\documentclass[aps,superscriptaddress,amsmath,amssymb,floatfix,twocolumn,showpacs,amsfonts,longbibliography,pra]{revtex4-1}
\usepackage{times}
\usepackage[varg]{txfonts}
\usepackage{textcomp}
\usepackage{graphicx}
\usepackage{subfigure}
\usepackage{tabu}
\usepackage{color}
\usepackage[colorlinks=true,citecolor=blue,urlcolor=blue,linkcolor=blue,hyperindex]{hyperref}
\usepackage{braket}
\usepackage{float}
\usepackage{overpic}
\usepackage{gensymb}
\usepackage{mathrsfs}
\usepackage{tikz}
\usepackage{bm}
\usepackage{multirow}
\usepackage{amsfonts}
\usepackage{amsmath}
\usepackage{mathrsfs}
\usepackage{amssymb}
\usepackage{paralist}
\usepackage{indentfirst}
\usepackage{ulem}
\usepackage{soul}
\usepackage{cancel}
\usepackage{CJK}

\allowdisplaybreaks

\begin{document}
\title{Spinon Singlet Pairing: Microscopic nature of plaquettes in stripy LDOS}
%\title{$4a_0$ periodicity and $2a_0$ shift of LDOS in a half-filled stripe}

\author{Ying Liang}
\affiliation{Department of Physics, Chongqing University, Chongqing, 401331, China}
\affiliation{Chongqing Key Laboratory for Strongly Coupled Physics, Chongqing University, Chongqing, 401331, China}

\author{Yi-Da Chu}
\affiliation{Beijing Computational Science Research Center, Beijing 100193, China}

\author{Hao Ding}
\affiliation{Department of Physics, Chongqing University, Chongqing, 401331, China}
\affiliation{Center of Quantum Materials and Devices, Chongqing University, Chongqing 401331, China}

\author{Shi-Jie Hu}
\thanks{Contact author: shijiehu@csrc.ac.cn}
\affiliation{Beijing Computational Science Research Center, Beijing 100193, China}
\affiliation{Department of Physics, Beijing Normal University, Beijing, 100875, China}

\author{Xue-Feng Zhang}
\thanks{Contact author: zhangxf@cqu.edu.cn}
\affiliation{Department of Physics, Chongqing University, Chongqing, 401331, China}
\affiliation{Chongqing Key Laboratory for Strongly Coupled Physics, Chongqing University, Chongqing, 401331, China}
\affiliation{Center of Quantum Materials and Devices, Chongqing University, Chongqing 401331, China}

\begin{abstract}
Scanning tunneling microscopy (STM) is a powerful tool for visualizing the local density of states (LDOS) of individual stripes in cuprates. However, the microscopic nature of the observed exotic LDOS patterns and their connection to high-$T_c$ superconductivity remain open questions.
Within the framework of the quantum colored string model, we reveal that the ubiquitous $4a_0\times4a_0$ plaquettes originate from either the breaking of local spinon singlet pairs through hole insertion, or the formation of an unpaired spinon upon electron addition in a stripe.
Moreover, by comparing our data with LDOS of cuprates, we identify an effect of particle-hole symmetry breaking (PHSB): a $2a_0$ shift, which is predicted and confirmed in a longer stripe ($L=18$).
At last, we establish and verify a general relation between hole density and plaquette size across multiple fillings.
Our work offers a fresh wavefunction-based perspective on interpreting STM signals in cuprate experiments and demonstrates that their origin may arise from spinon singlet pairing in the ground state of fluctuating stripes, the same mechanism underlying the $d$-wave sign structure [Phys. Rev. Lett. \textbf{137}, 086702 (2026)].
\end{abstract}
\maketitle

\textit{Introduction}.---The mechanism of superconductivity in high-$T_c$ cuprates stands as one of the most significant unsolved problems in physics over the past few decades~\cite{Zaanen_2023, HighTc1, HighTc2}.
In the underdoped regime, the interplay between stripes and $d$-wave superconductivity induces a complex and rich phase diagram~\cite{rmp_Kivelson, Uchida, dwave&stripe,yayu3,yayu4}. Recent advancements in STM have revealed complex stripy structures composed of a fundamental $4a_0\times4a_0$ plaquette ($a_0$ is the lattice constant of CuO$_2$ plane) that contains two holes~\cite{Kohsaka_2008, yayudwavepair, Wen4a0, yayuZRsinglet, yayustripe, halfB} and manifests three-nematic modulation~\cite{Wen4a0} with three bright bars (hereafter referred to as three-bar).
These in-situ experimental detections indicate that the pairing of superconducting quasiparticles may occur initially within a small local region~\cite{localpair1, localpair2, yayudwavepair}. On the other hand, numerical simulations up to date have demonstrated that the $\pi$-phase shift partially filled stripe neighbored by antiferromagnetic (AFM) domains~\cite{Tranquada_1995, Wen4a0} plays a critical role in building long-range $d$-wave superconductivity correlation in the ground states of the $t$-$J$~\cite{t-J1,t-J_ext2, tj_jiang} and $t$-$t'$-$U$ Hubbard models~\cite{t-J_ext1, t-J_ext3, dwave&stripe}.
Recently, density matrix renormalization group (DMRG) simulations successfully reproduced the LDOS of a short stripe with two $4a_0\times4a_0$ plaquettes~\cite{Yangshuo2025}, when the effective doping is close to $1/8$~\cite{note1}.
Moreover, these $4a_0\times4a_0$ plaquettes and other LDOS patterns are also observed throughout the underdoped region~\cite{yayuZRsinglet, yayustripe}.
However, their microscopic origin and connection to superconductivity remain elusive, leaving unclear whether they constitute essential signatures of high-$T_c$ superconductivity in STM experiments.

\begin{figure}[t]
	\centering
	\includegraphics[width=0.99\linewidth]{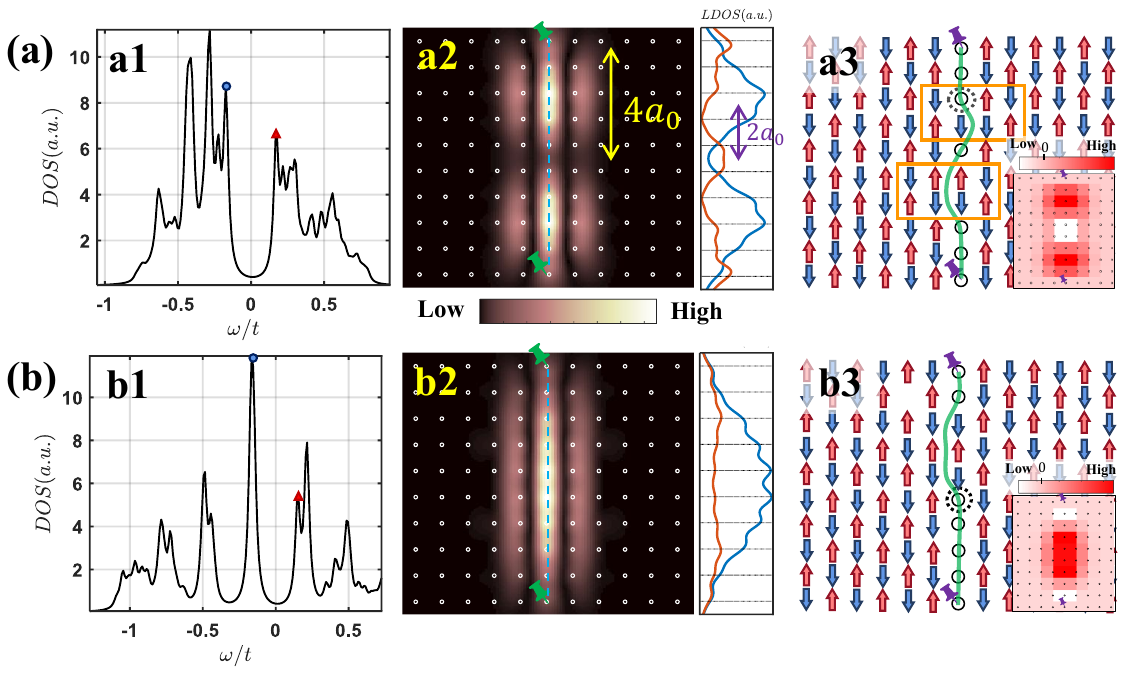}
    \caption{
     (a1, b1) The DOS \textit{vs}. energy $\omega/t$, where the states nearest to the Fermi level are indicated by peaks {\color{blue}$\bullet$} and {\color{red} $\blacktriangle$}).
    (a2, b2) The LDOS map for the state at peaks {\color{blue}$\bullet$} in an $L=10$ QCS. The endpoints of QCS are pinned (green tacks). The LDOS distribution for states at peaks {\color{blue}$\bullet$} (blue line) and {\color{red} $\blacktriangle$} (red line) along the stripe $y$ mirror axis (blue dashed line) is shown on the right side.
    (a3, b3) The largest-weight basis in $\ket{g^{N+1}}$. Inset: the deviation in spinon density distribution $\Delta \rho_\textbf{r}(i_x,\,i_y)$. $J/t=0.6$ is held fixed throughout. In \textbf{(b)}, the spin exchange interaction strength $J_{xy}$ is turned off as a controlled tuning to test the stability of the $4a_0\times4a_0$ plaquettes.}
    \label{fig1}
\end{figure}

%The QCSM
Recently, the quantum colored string (QCS) model (QCSM) has been proposed as a quantitative effective description of the low-energy physics of the hole-doped fermionic models~\cite{QCSM1, QCSM2}, in which the two-dimensional ($2$D) electron configuration is encoded through exotic quasi-particles (e.g., spinons and holons) that self-organize into one-dimensional ($1$D) stripes. Not only has its quantitative validity been benchmarked through comparison with DMRG results, but the QCSM can also unveil that the local $d$-wave sign structure directly results from the pairing of the spinons~\cite{QCSM2}. Therefore, it is straightforward to ask whether these characteristic features in the stripe LDOS also originate from spinon singlet pairs?

%The summary part (modified at last)
In this Letter, we calculate the LDOS for a long stripe within the QCSM framework. The numerical results capture the main features observed in STM experiments, particularly the $4a_0 \times 4a_0$ plaquette. As demonstrated in Fig.~\ref{fig1}(a),  we find the $4a_0$ period originates from the spinon singlet pair. After elucidating its microscopic mechanism, %which is relevant to spinon singlet pairs, 
we identify a $2 a_0$ shift and confirm it in a larger system ($L=18$), which has also been observed in experiments~\cite{yayuZRsinglet}. The spin exchange interaction strongly benefits the pairing of the spinon, and we find that its switching off eliminates not only the $d$-wave pattern but also the $4a_0 \times 4a_0$ plaquette [see Fig.~\ref{fig1}(b)]. Finally, we extend our analysis to general filling and quantitatively show the relationship between the fillings and the plaquette period. Distinct from previous work~\cite{Yangshuo2025}, no impurity potential is introduced here, which indicates that the $4a_0 \times 4a_0$ plaquette directly originates from the spinon singlet pairing, rather than disorder-induced inhomogeneity. Therefore, the number of plaquettes at negative bias is exactly the same as the number of pairings.

\begin{figure}[t]
    \centering
    \includegraphics[width=0.99\linewidth]{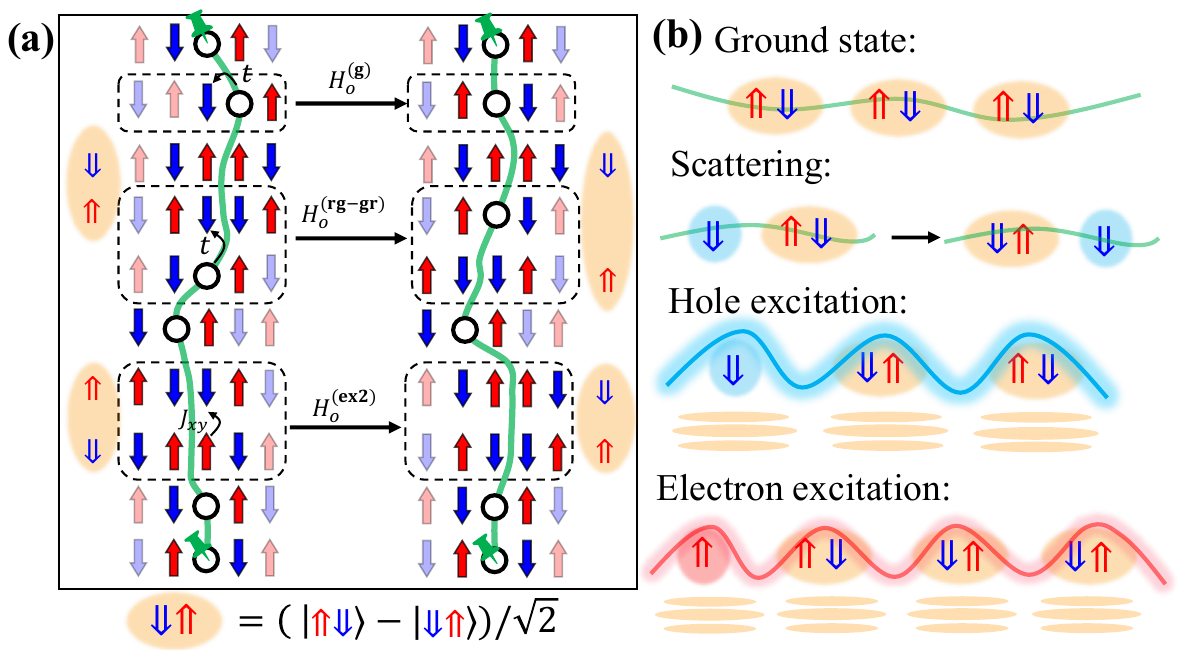}
    \caption{
     (a) The schematic illustrations for a half-filled QCS (green lines) with $L=10$. Spinon singlets (orange ellipses) and the most relevant processes defined in $H_\text{e}^\text{QCS}$ (Eq.~\eqref{Eq1}) (enclosed in black dashed boxes) are highlighted.
    (b) The schematic for the relation between spinon scattering and the number of plaquettes in the LDOS map. Blue (Red) line: the standing wave of spinons in the hole (electron) excited state. The orange shadows below are the LDOS patterns generated by the scattered spinons.}
    \label{fig2}
\end{figure}

% Introduce the model and method
%Model
\textit{Model and Methods}.---In the underdoped region, it is commonly believed that spin-charge separation occurs within a two-dimensional AFM background.
Consequently, holons and spinons often self-organize into a stripe pattern, each of which manifests as a unidirectional, one-dimensional topological defect characterized by a $\pi$-phase shift~\cite{stripe_pi_phase} with a finite length $L$, observed in experiments~\cite{yayustripe, yayuZRsinglet, Wen4a0}.
The main configurations of a single stripe can be encoded into the QCS representation as shown in Fig.~\ref{fig2}(a): $|\chi_\text{s}, \{\Gamma^{z}\}, \{c_{\chi}\}\rangle$, which includes three types of color quasi-particles (CQPs) $\{c_{\chi}\}$:  spinons (color $c_{\chi}=\textbf{r}$, e.g., $\uparrow$$\downarrow$$\downarrow$$\uparrow$), holons (color \textbf{g}, e.g., $\uparrow$$\circ$$\downarrow$), and dual-holes (color \textbf{b}, e.g., $\uparrow$$\circ$$\circ$$\uparrow$)~\cite{QCSM1, QCSM2}.
The relative distances along the $x$-axis between CQPs in neighboring rows are denoted with the effective spin fields (ESFs) $\{\Gamma^{z}\}$ ($\Gamma^z_{\bar{y}}/a_0 \in \mathbb{Z}$)~\cite{QCSM1, QCSM2}.
Two fixed endpoints, pinned by external fields, have a relative shift $\chi_\text{s}$ along the $x$-axis.
Furthermore, there is no constraint in the $x$ direction, so the system can be taken as a square lattice with $(L_x, L_y)=(+\infty, L)$.
In the QCS representation, the $t$-$J$ model can be effectively derived as follows:
\begin{eqnarray}
\begin{split}
    % H_{e}^{CS} &= J_z H_{d} -t\left( H_{o}^{(g)} + H_{o}^{(rb-gg)} + H_{o}^{(rg-gr)} + H_{o}^{(bg-gb)} \right) \nonumber \\
    % &+ J_{xy}\left(H_{o}^{(ex1)} + H_{o}^{(ex2)}\right),\\
    H_\text{e}^\text{QCS} &={H_\text{d}(J)} - H_\text{o}^{(\textbf{g})} (t)+ H_\text{o}^{(\textbf{rb}-\textbf{gg})}(t) + H_\text{o}^{(\textbf{rg}-\textbf{gr})} (t) \\
    &+ H_\text{o}^{(\textbf{bg}-\textbf{gb})}(t)+ H_\text{o}^{(\textbf{ex1})} (J_{xy})+ H_\text{o}^{(\textbf{ex2})}(J_{xy})\,.
    \label{Eq1}
\end{split}
\end{eqnarray}
The diagonal term $H_\text{d}$ represents the tension energy of QCS, which is proportional to $|\Gamma^z|$ and the Ising interaction strength $J$.
The hopping coefficient $t$ governs all hopping terms, including $H_\text{o}^{(\textbf{g})} $ for holon hopping along the $x$-axis, $H_\text{o}^{(\textbf{rb}-\textbf{gg})} $ for the creation and annihilation of spinon and dual-hole from a pair of neighboring holons, and $H_\text{o}^{(\textbf{rg}-\textbf{gr})}$ and $ H_\text{o}^{(\textbf{bg}-\textbf{gb})}$ for the movement of spinons and dual-holes along QCS, respectively.
The spin exchange interaction strength $J_{xy}=2 J$ facilitates spinon pairing through two processes $H_\text{o}^{(\textbf{ex1})}$ and $H_\text{o}^{(\textbf{ex2})}$. They exchange chirality ($\textbf{r}_{\Uparrow}:\uparrow\downarrow\downarrow\uparrow$ and $\textbf{r}_{\Downarrow}:\downarrow\uparrow\uparrow\downarrow$), thereby forming a singlet denoted by $\st{\textbf{r}\textbf{r}}=(\ket{\textbf{r}_{\Uparrow}\textbf{r}_{\Downarrow}}-\ket{\textbf{r}_{\Downarrow}\textbf{r}_{\Uparrow}})/\sqrt{2}$ to further reduce the energy.
Those terms in Eq.~\eqref{Eq1} make the leading order contribution to the ground state wave function are shown in Fig.~\ref{fig2}(a), and the other higher order fluctuation leads to the spinon singlet coherence and $d$-wave patterns.
The details can be found in {Refs}.~\cite{QCSM1, QCSM2}.

%Method
In this work, we consider a QCS with $\chi_\text{s}=0$ and $1$, and impose that the ground state $\ket{g^N}$ contains $N=L/2+1$ holes.
Consequently, the fluctuating $(L-2)$ rows of the QCS have $(L-2)/2$ holes, excluding the two pinned endpoints.
This setting effectively realizes a half-filled stripe that is strongly entangled with $d$-wave superconductivity in the cuprates.
For simplicity, we henceforth refer to this stripe as ``a half-filled stripe" or ``a half-filled QCS".
In addition, the single-particle excited states $\ket{m^{N\pm1}}$ exist within the Hilbert space, each containing $N\pm1$ holes.
Among these, the two single-particle excited states corresponding to the first peaks in the DOS just below and above the Fermi level (i.e., energy $\omega=0$) are labeled as $\ket{g^{N\pm1}}$.
The ESF has a cutoff of $\Gamma^z_\text{max}=6a_0$, ensuring that $|\Gamma^z| \leq \Gamma^z_\text{max}$.
This cutoff permits a large width of the QCS along the $x$-axis, accommodating a potential region of string fluctuations.
After getting $\ket{g^N}$ and $\ket{g^{N\pm1}}$, we can compute the LDOS [see details in End matter A].
To guarantee that the half-filled stripe exhibits $d$-wave pairing patterns, we typically choose $t=1$ (as energy unit), $J=0.6$, and double $J_{xy}$ to account for renormalization effects induced by the AFM background~\cite{QCSM2}.

%Short stripe
\textit{Stripe and $4a_0$ period}.---We first consider a short half-filled QCS with $L=10$ [Fig.~\ref{fig1}]. When both QCS endpoints are aligned, i.e., $\chi_\text{s}=0$ [Fig.~\ref{fig1}(a)], the LDOS map for the first peak below the Fermi level (corresponding to $\ket{g^{N+1}}$) shows a stripy pattern with a period of $4a_0$ along the $y$-axis.
Each $4a_0\times4a_0$ plaquette features a three-bar pattern, with the middle bar brighter and the side bars slightly darker, forming a $D_2$ symmetry [Fig.~\ref{fig1}(a2)].

%The spinon snapshot %spinon distribution
To diagnose the microscopic nature of the $4a_0\times4a_0$ plaquette, we first evaluate the importance of the basis in Fock space, utilizing the amplitudes of the wave function coefficients~\cite{QCSM2}.
The most important basis of the ground state $\ket{g^N}$ for the half-filled stripe reveals two separate spinon pairs, which are crucial for $d$-wave sign structure~\cite{QCSM2}.
In contrast, the largest-weight basis of the single-particle excited state $\ket{g^{N+1}}$ [Fig.~\ref{fig1}(a3)] shows that one of the two spinon pairs (orange boxes) is disrupted by removing a spin-up electron (black dashed circle), resulting in the emergence of an unpaired spinon.
Moreover, we introduce the deviation $\Delta\rho_\textbf{r}(i_x,\,i_y) = \rho_\textbf{r}^{N}(i_x,\,i_y) -\rho_\textbf{r}^{N+1}(i_x,\,i_y)$, which quantifies the difference between the spinon density distributions $\rho_\textbf{r}^{N}(i_x,\,i_y)$ and $\rho_\textbf{r}^{N+1}(i_x,\,i_y)$ for the states $\ket{g^N}$ and $\ket{g^{N+1}}$.
The $2$D lattice site $i$ is labeled by $i_x$ and $i_y$ representing its coordinates along the $x$ and $y$ axes, respectively.
As demonstrated in the inset of Fig.~\ref{fig1}(a3), the strongest reduction in spinon density occurs precisely where the spin-up electron has been removed.
These findings suggest that each $4a_0\times4a_0$ plaquette in $\ket{g^N}$ is composed of a spinon singlet pair and two holons, with their movement along the QCS giving rise to those exotic LDOS patterns.
Furthermore, the pairing gap is $U$-shaped and approximately $0.35t$ across the entire QCS with PHSB[Fig.~\ref{fig1}(a1)], which coexists with $d$-wave sign structure [see Supplemental Material (SM) and Ref.~\cite{QCSM2}] and keeps consistent with the $dI/dV$ spectra of dilute stripes observed in STM experiment~\cite{yayuZRsinglet}.

Previous studies~\cite{QCSM2} have shown that the formation of the spinon pairs strongly depends on the spin exchange interaction.
Therefore, it can be inferred that switching it off may ruin the $4a_0\times 4a_0$ plaquettes.
To check it, we investigate a half-filled QCS for the case of $J_{xy}=0$, while keeping all other parameters consistent with those used in Fig.~\ref{fig1}(a). 
The LDOS map for the state $\ket{g^{N+1}}$ [left panel in Fig.~\ref{fig1}(b2)] shows a continuous three-bar pattern that lacks the $4a_0$ periodicity. This finding is further supported by the LDOS distribution along the $y$ mirror axis of the stripe [the right panel of Fig.~\ref{fig1}(b2)], where no nodal points with low LDOS values are observed in the vicinity of the central region.
Additionally, the largest-weight basis of the state $\ket{g^{N+1}}$ shows that 
all the spinons are clustered together [Fig.~\ref{fig1}(b3)], which indicates the spinon singlet pairs are not energetically favored. 
This scenario is corroborated by the deviation of the spinon density distribution $\Delta\rho_\textbf{r}$ shown in the inset of Fig.~\ref{fig1}(b3). On the other hand, turning off the spin exchange interaction also destroys the $d$-wave correlation (see SM for details). 
Therefore, we think that the $4a_0\times4a_0$ pattern can be taken as the sign of the local pairing related to the d-wave superconductivity correlation, but not enough as the smoking gun. Meanwhile, our success in reproducing the experimental phenomenon in a single QCS indicates that the $4a_0\times4a_0$ plaquettes are not independent units~\cite{yayudwavepair}, but rather emerge from the inherent structure of the fluctuating QCS with a $\pi$-phase shift. Furthermore, the ladder pattern at a large energy bias can also be captured by the QCSM, and the plaquette widening is found to be related to the quantum fluctuation of the QCS [see details in End matter B].

%hole & electron
Here, we can conclude the microscopic mechanism of the $4a_0\times4a_0$ plaquettes in a half-filled stripe. When a hole is inserted into the ground state $\ket{{g}^N}$ for a half-filled stripe of finite lengths, one of the spinon pairs is disrupted to form a single spinon (e.g., $\ket{...\st{\textbf{r}\textbf{r}}.. \st{\textbf{r}\textbf{r}}...}\rightarrow  \ket{...\textbf{g}\textbf{r}_\Uparrow  ..\st{\textbf{r}\textbf{r}}...}$),  as reflected in the largest-weight basis of the state $\ket{{g}^{N+1}}$ [Fig.~\ref{fig1}(a3)] and the schematic shown in Fig.~\ref{fig2}(b). In the ground state $\ket{g^N}$, there are $N_\mathrm{sp}=(L-2)/4$ spinon pairs, allowing the hole insertion to occur in $N_\mathrm{sp}$ possible positions.
Therefore, we can find the number of $4a_0\times4a_0$ plaquettes is the same as the number of spinon pairs $N_\mathrm{sp}$ as shown in Fig.~\ref{fig1}(a2).
Differently, when an electron is inserted, it does not destroy the original spinon singlet pairs but instead creates a new single spinon in the state $\ket{g^{N-1}}$, thus yielding $(N_\mathrm{sp}+1)$ possible positions in total for placing the generated single spinon.
As a result, we expect $(N_\mathrm{sp}+1)$ plaquettes in the LDOS map. %[Figs.~\ref{fig3}(d) and (e)], where the middle plaquettes retain the $4a_0\times4a_0$ structure, while those near the edges do not strictly conform to this arrangement.
Since the LDOS map of the electron excitation has one more plaquette than that of the hole excitation, a half-period ($2a_0$) shift must be introduced to accommodate the constraints imposed by the finite stripe length.
Notably, through the splitting-pairing process $\st{\textbf{r}\textbf{r}}\leftrightarrow$ $\textbf{r}_\Downarrow \textbf{r}_\Uparrow$ or $\textbf{r}_\Uparrow\textbf{r}_\Downarrow$, this single spinon can propagate within the stripe and generate new configurations, e.g., $\ket{...\textbf{r}_\Uparrow  ..\st{\textbf{r}\textbf{r}}...}\rightarrow \ket{...\st{\textbf{r}\textbf{r}}..\textbf{r}_\Uparrow  ...}$ as shown in Fig.~\ref{fig2}(b).

\begin{figure}[t]
    \centering
    \includegraphics[width=0.99\linewidth]{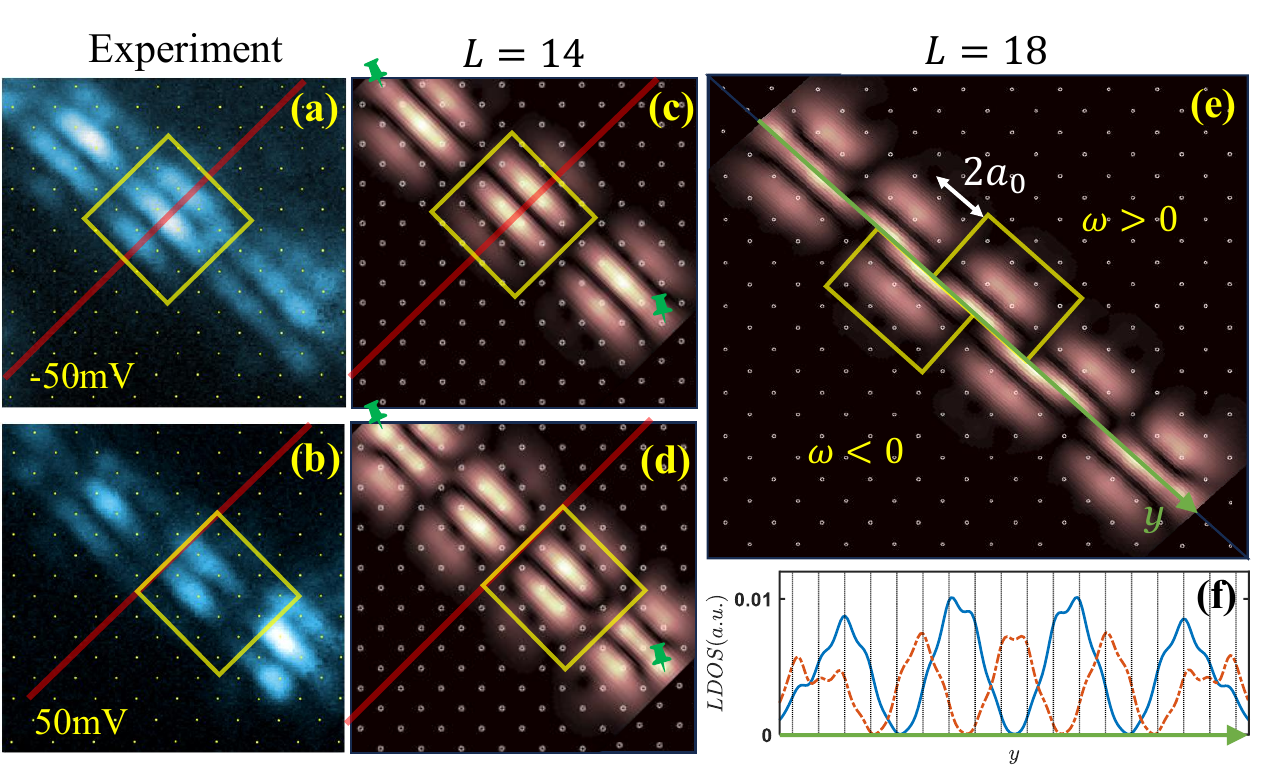}
    \caption{(a,b) Stripy patterns in the LDOS map observed in hole-doped Ca$_2$CuO$_2$Cl$_2$ (doping $p=0.03$) with an energy bias of $\pm50$mV, as presented in Figs.~2(d,e) of Ref.~\cite{yayuZRsinglet}.
    The DMRG-calculated LDOS map at $L=14$  with $\chi_s=1$ is shown for (c) $\ket{g^{N+1}}$ below the Fermi level and for (d) $\ket{g^{N-1}}$ above the Fermi level.
    The yellow boxes highlight the $4a_0\times4a_0$ structure, accompanied by the same red reference line.
    (e) The LDOS map of a longer stripe ($L=18$) with $\chi_s=0$, displayed below (left-lower) and above (right-upper) the Fermi level.
    (f) The LDOS along a green arrow in (e) for $\ket{g^{N+1}}$ (blue line) and $\ket{g^{N-1}}$ (red line), with black lines indicating Cu atom positions.}
    \label{fig3}
\end{figure}
%PHS Breaking and two-bar
\textit{$2a_0$ shift.---}
Most intriguingly, a novel $2a_0$ shift of the $4a_0$ periodic structure [Figs.~\ref{fig3}(a,b)] has been detected, yet it is unnoted within the STM data for cuprates at ultra-low doping $p=0.03$~\cite{yayuZRsinglet}, where the stripe is longer than $L=10$.
Thus, we further employ the recently developed DMRG method to simulate the LDOS of a longer QCS~\cite{DMRG_QCSM} [see details in SM].
For a QCS with $L = 14$ ($N_\mathrm{sp}=3$), the LDOS map presents a pattern {[Figs.~\ref{fig3}(c,d)]} similar to the experimental results {[Figs.~\ref{fig3}(a,b)]}.
As discussed above, we can observe that the electron-excited LDOS map has one more $4a_0\times4a_0$ plaquette compared with the hole-excited case, so a $2 a_0$ shift between positive and negative bias can be clearly identified.
Further, we simulate a larger $L = 18$ ($N_\mathrm{sp}=4$) [Fig.~\ref{fig3}(e)] and extract the LDOS of the stripe center section [Fig.~\ref{fig3}(f)].
The simulation results indicate that the $2a_0$ shift is very robust as the stripe length increases.
% two-bar
We also noticed that in the stripes observed in the STM experiments, apart from the three-bar shape, the two-bar shape was also quite common. And this stripe can be achieved by misaligning its two endpoints [Fig.~\ref{fig3}(c, d)].
% In contrast, when the two endpoints of the QCS are misaligned by $a_0$ [Fig.~\ref{fig1}(c)], the $4a_0$ periodicity persists; however, the previously mentioned $D_2$ symmetry of the brightness distribution in the stripy pattern no longer holds.
% Specifically, the three-bar pattern in the $4a_0\times4a_0$ plaquette is replaced with a two-bar pattern [Fig.~\ref{fig1}(c2)], where the leftmost bar on the bottom plaquette and the rightmost bar on the upper plaquette appear relatively darker.
% Therefore, the two-bar pattern commonly observed in experiments~\cite{yayudwavepair, Wen4a0, yayuZRsinglet, yayustripe, halfB} likely arises from this misalignment, where the two endpoints of a stripe may be pinned by disordered potentials.
% Furthermore, the misalignment of the endpoints does not significantly affect the DOS.

\begin{figure}[t]
    \centering
    \includegraphics[width=0.99\linewidth]{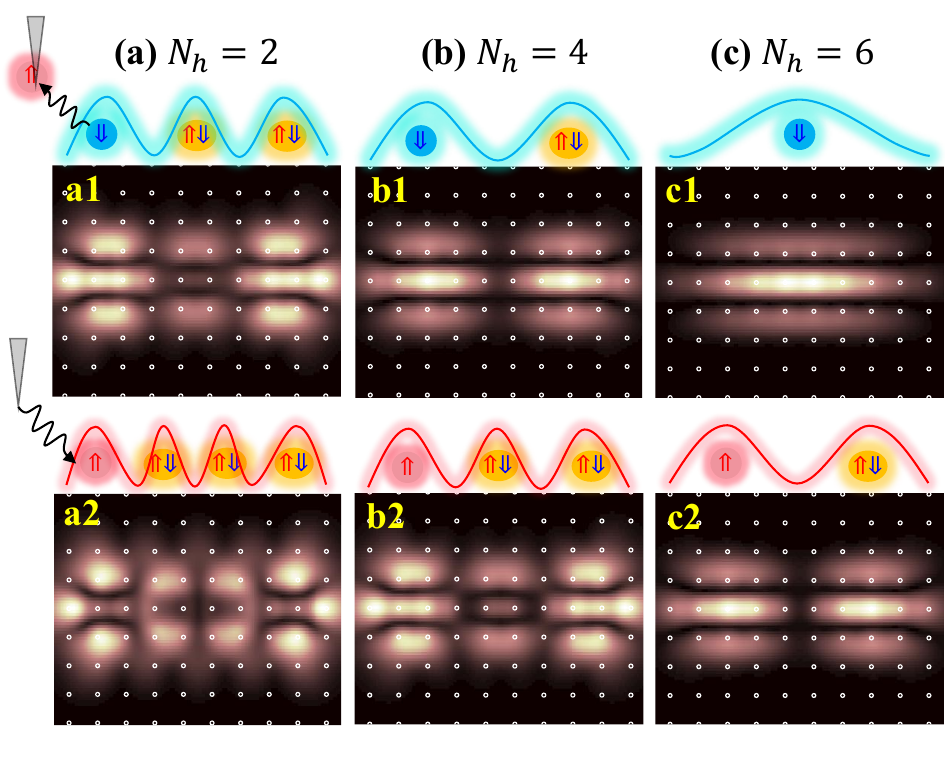}
    \caption{The LDOS maps of a $L=10$ stripe with free hole number (a) $N_h=2$, (b) $N_h=4$, (c) $N_h=6$. Top row: Patterns of hole excitation; bottom row: Patterns of electron excitation. The solid, glowing lines in the illustration represent the standing wave of a single spinon, which gives rise to the LDOS patterns shown below.}
    \label{fig4}
\end{figure}

Actually, such a $2a_0$ shift with PHSB can be extended to the general filling case. As shown in Fig.~\ref{fig4}, we fixed the length $L$ of the stripe and adjusted the number of holes $N$ within the stripe. Removing the fixed ends of holons, the number of free holes is $N_h=N - 2$. Accordingly, the number of spinon singlets is $N_\mathrm{sp}=(L-2-N_h)/2$, and the plaquette number naturally changes with $N_h$. If we define the length of the free part of the stripe as $L_f=L-2$ and the hole density as $\rho_h=N_h/L_f$. Then, for the case of hole excitation, the length (or period) $L_p^{(h)}$ of the plaquette along the stripe direction satisfies
\begin{equation}
    \frac{1}{L_p^{(h)}}=\frac{\rho_h}{2},
\end{equation}
and for the case of electron excitation, it is
\begin{equation}
    \frac{1}{L_p^{(e)}}=\frac{1}{L_p^{(h)}}+\frac{1}{L_f}.
    \label{Eqn3}
\end{equation}
Thus, the $4a_0$ period with $2a_0$ shift is actually a special case of a finite-length stripe with $\rho_h=0.5$. Conversely, the LDOS period can be used to infer the stripe filling. 

From Eq.~\eqref{Eqn3}, we can find that the difference between electron and hole excited cases is strongly suppressed in a long stripe. However, in the extremely low doping case \cite{yayuZRsinglet}, the length of the stripe is usually very short, so the PHSB becomes more serious. Therefore, the number of spinon singlet pairs can only be obtained by counting the plaquettes for the hole-excited LDOS map (negative energy bias). 

\textit{Conclusion and discussion}.---The LDOS map of a long half-filled stripe is analyzed via the QCSM, which shows strong agreement with many experimental observations.
Through a wavefunction-based analysis, the microscopic nature of the $4a_0\times 4a_0$ plaquette in the LDOS maps is revealed, closely linked to the interplay between spinon singlet pairs and an inserted hole or electron.
Given the importance of spinon singlet pairs for $d$-wave pairing~\cite{QCSM2}, we concur with the experimental physicists in considering the $4a_0\times 4a_0$ plaquette as a precursor to the ``Cooper pair".
Furthermore, the short-range spinon singlet pairing is possibly ubiquitous and opens a local spin singlet excitation gap in the pseudogap regime, which is implied in the high-field NMR experiment~\cite{twogap}.
Most critically, we discover and confirm a $2a_0$ shift between positive and negative energy bias in a long stripe ($L=18$).
This phenomenon appears to be ubiquitous in the underdoped region of the cuprates, and we believe it may be even more general.

By comparing the QCSM and molecular orbital pictures~\cite{yayuZRsinglet}, we find that both effectively capture the fundamental physics of the $4a_0\times 4a_0$ plaquette: local pairing and two-hole filling.
However, the QCSM offers a more comprehensive understanding of the underlying microscopic mechanisms.
{Upon hole doping away} from a perfect AFM insulator, the $4a_0\times 4a_0$ plaquettes initially tend to form a line structure rather than being arranged randomly.
Moreover, the QCS, as a one-dimensional topological defect, is compatible with several key features simultaneously, including the $\pi$-phase shift, fractional excitations, $d$-wave superconducting correlations, $4a_0$ periodicity, and local pairing.
Apart from the pinned stripe endpoints, our analysis using the QCSM introduces no impurity effects, suggesting that the $4a_0$ periodicity and $2a_0$ shift are intrinsic rather than being induced by broad impurity potential fields~\cite{Yangshuo2025}.
However, eliminating the influence of impurities in the experiment remains a significant challenge. The C$_4$ symmetry breaking in the ARPES measurement of disorder-removed CuO$_2$ plane has implies the existence of large-scale unidirectional stripes~\cite{ARPES_C4break}, while the realization of in-situ observation can be achieved by regulating the local hole concentration through gate voltage to eliminate the influence of impurity potential.
Definitely, the validity of our scenario requires further elaborate experiments and numerical simulations conducted on longer stripes even for other periods with different hole fillings.

%future work
%\change{Our existing results provide a pictorial and concrete playground for quantitatively analyzing the microscopic mechanism of the LDOS of the long finite-size stripe. More importantly, the interactions among the multi-stripes under study may be the key to establishing global phase coherence, which may be reflected in the disappearance of the $2a_0$ shift due to the release of the spinon scattering dimensional restrictions. We believe that our results and future extensions will gradually establish a bridge between high-$T_c$ superconductivity and its microscopic nature. Furthermore, it may inject new vitality into the understanding of strong correlation phenomena from the perspective of high-dimensional topological objects.}{}

We would like to thank Hui-Xia Fu and Yayu Wang for many helpful discussions. This work is supported by grants: MOST 2022YFA1402700, NSFC 12274046, NSFC 12547101, and Xiaomi Foundation / Xiaomi Young Talents Program. Computational resources from Tianhe-2JK at the Beijing Computational Science Research Center are also highly appreciated.

\bibliography{ref}

\section*{End Matter}
\subsection{Method for LDOS calculations}
%LDOS (Parts of them may be moved to endnote)
To explore the LDOS of the half-filled stripe with $N$ holes, we adopt the large-scale ED to obtain the ground state $\ket{g^N}$ as well as a hundred low-energy excited states ($50$ for $\ket{m^{N+1}}$ and $50$ for $\ket{m^{N-1}}$) for an individual stripe defined in the main text, achieving a scale comparable to that used in the DMRG simulations~\cite{Yangshuo2025}.
Recently, we have also developed the DMRG method for simulating QCSM in larger stripes ($L\ge14$) (see details in SM).
Then, to connect with spectroscopic measurements, we calculate the zero-temperature single-particle spectral function $A(i, j, \omega)$, which is defined as
%Its general definition in real space involves the $N$-electron ground state $|g^N\rangle$ and the low-lying states $|m^{N\pm1}\rangle$ of the $(N\pm1)$-electron system:
\begin{eqnarray}
\begin{split}
    A(i, j, \omega)\!=\!&\sum_{m,\sigma} \braket{g^N \vert c^{\phantom{\dag}}_{i,\sigma} \vert m^{N-1}} \braket{m^{N-1} \vert c_{j,\sigma}^\dagger \vert g^N} \delta (\omega\!-\!E_m^{N-1}\!+\!E_g^N) \nonumber \\
    + &\sum_{m,\sigma} \braket{g^N \vert c_{j,\sigma}^\dagger \vert m^{N+1}} \braket{m^{N+1} \vert c^{\phantom{\dag}}_{i,\sigma} \vert g^N} \delta(\omega\!-\!E_g^N\!+\!E_m^{N+1})\,,
    \label{Eq2}
\end{split}
\end{eqnarray}
where $c^{\phantom{\dag}}_{i,\sigma}(c_{i,\sigma}^{\dagger})$ is the annihilation (creation) operator of an electron with spin $\sigma=\uparrow,\downarrow$ at site $i$. For direct comparison with the differential conductance from STM, we compute the LDOS map in the $2$D plane by superimposing Wannier orbitals~\cite{wannier1,wannier2}.
The spectral function $\tilde{A}(\boldsymbol{d}, \boldsymbol{d}',\omega)$ in real space is constructed from the lattice spectral function $A(i, j, \omega)$ as
\begin{equation}
\tilde{A}(\boldsymbol{d}, \boldsymbol{d}', \omega) = \sum_{i, j} w^{\phantom{\dag}}_i(\boldsymbol{d}) w_j^{*}(\boldsymbol{d}') A(i,j,\omega)\,,
\label{Eq3}
\end{equation}
where the Wannier function $w_i(\boldsymbol{d})$ is centered at the $i$-th Cu site, and $\mathbf{d}=(d_x, d_y)$ gives the $2$D coordinates in the real space.
Consequently, the LDOS is given by $\tilde{A}(\boldsymbol{d}, \boldsymbol{d}, \omega)$.

%ladder pattern
\subsection{Ladder pattern}
\begin{figure}[h]
    \centering
    \includegraphics[width=0.99\linewidth]{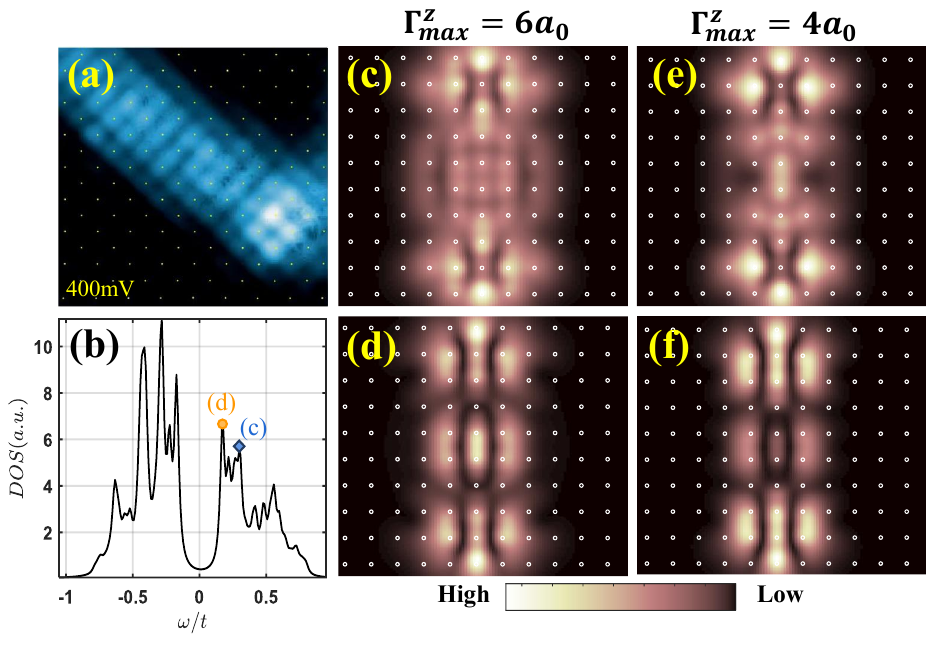}
    \caption{(a) The ladder pattern in the LDOS map observed in hole-doped Ca$_2$CuO$_2$Cl$_2$ (doping $p=0.03$) with an energy bias of $\pm400$mV, as presented in Fig.~2(g) of Ref.~\cite{yayuZRsinglet}.
    For a half-filled QCS with $L=10$, (b) the DOS \textit{vs}. $\omega$, and the LDOS maps for (d,f) the first peak and (c,f) another higher peak above the Fermi level are plotted.
    In (b), the first peak ({\color{orange}$\bullet$}) and the second peak ({\color{blue}$\blacklozenge$}) are highlighted.
    We set $\Gamma^z_\text{max}=6a_0$ for (b,c,d),  while $\Gamma^z_\text{max}=4a_0$ for (e,f).}
    \label{figs}
\end{figure}

For a large energy bias, the LDOS map exhibits a multi-rung ladder pattern in experiments~\cite{yayuZRsinglet}.
The microscopic mechanism underlying this phenomenon is expected to align with the framework of QCSM, particularly when considering the highly-excited states far above the Fermi level.
In Fig.~\ref{figs}(b), several peaks can be observed above the U-shaped gap in the DOS.
For positive bias ($\omega > 0$), the pinned holons at both stripe endpoints contribute to a four-leaf clover structure, which is consistent with previous experimental findings~\cite{yayuZRsinglet} and numerical simulations~\cite{Yangshuo2025}.
Furthermore, the LDOS map corresponding to the first peak (i.e., $\ket{g^{N-1}}$) [Fig.~\ref{figs}(d)] shows a typical stripe pattern with three $4a_0 \times 4a_0$ plaquettes, mirroring the scenario mentioned above.
In examining another peak with a higher energy [Fig.~\ref{figs}(c)], the central LDOS map reveals a ladder pattern.

% Interpretation

We further investigate how the patterns change in response to different values of the ESF cutoff $\Gamma^z_\text{max}$.
By comparing the LDOS maps at $\Gamma^z_\text{max} = 6a_0$ [Fig.~\ref{figs}(d)] and $\Gamma^z_\text{max} = 4a_0$ [Fig.~\ref{figs}(f)], we observe that the stripe pattern remains nearly unchanged.
In contrast, the width of the ladder pattern at $\Gamma^z_\text{max} = 6a_0$ [Fig.~\ref{figs}(c)] is significantly larger than that at $\Gamma^z_\text{max} = 4a_0$ [Fig.~\ref{figs}(e)].
This finding suggests that the vibration strengths of the stripe differ between the two patterns, and the ladder pattern likely arises from enhanced string fluctuations associated with larger $\Gamma^z_\text{max}$.
Thus, we argue that the increasing energy bias elevates the vibrational strength of the stripe, which can cause a transition of the LDOS map from a stripe pattern to a ladder pattern.

\newpage

\renewcommand{\thesubsection}{S\arabic{subsection}}
\setcounter{subsection}{0}

\renewcommand{\theequation}{S\arabic{equation}}
\setcounter{equation}{0}

\renewcommand\thefigure{S\arabic{figure}}
\setcounter{figure}{0}

\section*{Supplemental Material}

\section{$d$-wave pairing dependence on the spin exchange interaction}

\begin{figure}[h]
    \centering
    \includegraphics[width=0.99\linewidth]{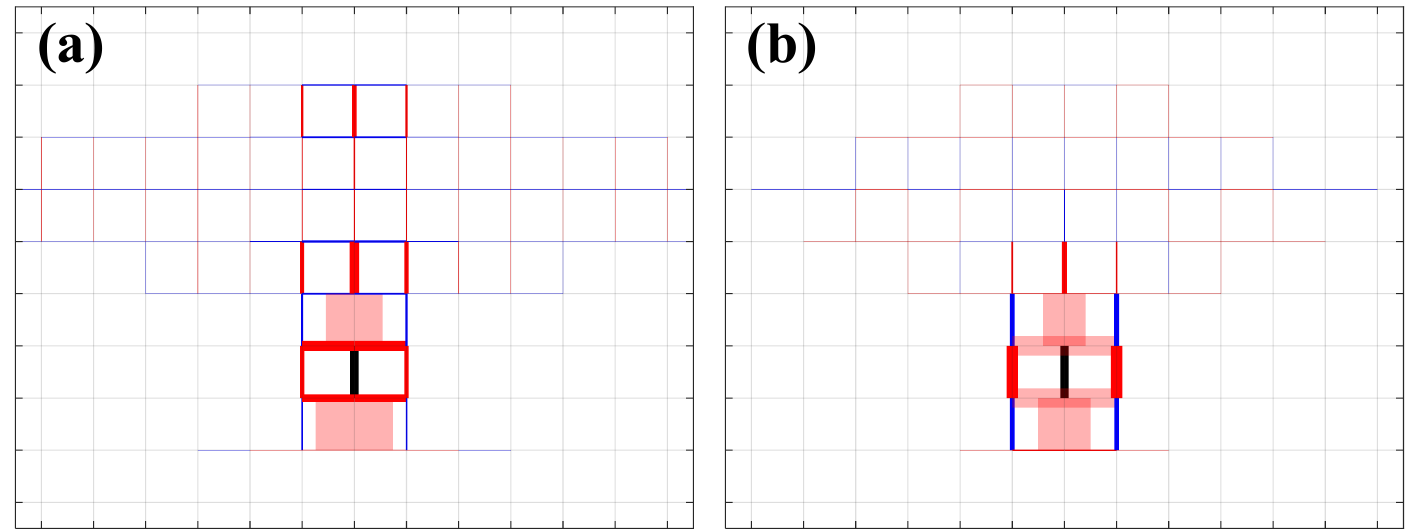}
    \caption{The PPC function $G_{b,b'}$ for $\ket{g^N}$ at (a) $J_{xy}=0.6$ and (b) $J_{xy}=0$ in a half-filled QCS. All other QCSM parameters are identical to those in Fig.~1(b) of the main text.}
    \label{figs3}
\end{figure}

The absence of spinon pairs leads not only to the disappearance of the $4a_0$ periodicity, as mentioned in the main text, but also to the disruption of the $d$-wave pairing patterns in the pair-pair correlation (PPC) function $G_{b,b'}=\braket{\Delta^\dagger_b \Delta^{\phantom{\dag}}_{b'}}$ between bonds $b$ and $b'$, owing to the lack of spinon singlet pair [Fig.~\ref{figs3}].

\section{DMRG calculations}
In Figs.~3(c,d,e,f) of the main text, we employ DMRG to simulate a half-filled stripe with lengths $L=14$ and $18$ as described by the QCSM.
In the QCSM, the relative distances along the $x$-axis between two CQPs in neighboring rows must adhere to a configuration-dependent gauge constraint, which is linked to the corresponding ESF value $\Gamma^z$.
For example, for CQPs $\mathbf{g}$ and $\mathbf{r}$, $\Gamma^z$ should always be odd numbers, whereas for $\mathbf{g}$ and $\mathbf{b}$, $\Gamma^z$ remains even.
Therefore, in the model definition part of the DMRG code, we introduce additional bonus terms for ensuring these constraints are satisfied, e.g., $-\lambda n^\mathbf{g}_y f_o (\Gamma^z_y) n^\mathbf{r}_{y+1}$ and $-\lambda n^\mathbf{g}_y f_e (\Gamma^z_y) n^\mathbf{b}_{y+1}$, using a large value of $\lambda=10^4$.
The selection function $f_o(x) = 1$ for odd $x$ and $f_o(x) = 0$ otherwise.
Similarly, $f_e(x) = 1$ for even $x$ and $f_e(x) = 0$ otherwise.
At the two stripe endpoints, we include a term of $h (n^\mathbf{r} + n^\mathbf{b})$ with a pinning field of $h=10^6$.
With a truncation dimension of $m=4096$, ensuring a convergence threshold of $10^{-7}$, we obtain the states $\ket{g^N}$, $\ket{g^{N+1}}$ and $\ket{g^{N-1}}$ by adjusting the number of holes.
Finally, we calculate the LDOS maps as detailed in End Matter A.

\end{document}